\documentclass[preprint,showpacs,showkeys]{revtex4}
\usepackage{graphicx}
\usepackage{dcolumn}
\usepackage{amsmath}

\begin{document}

\title {The energy spectrum of cosmic rays above $10^{15}$ eV as derived from
air Cherenkov light measurements in Yakutsk}

\author{A.A. Ivanov}\email[Corresponding author. E-mail: ]{a.a.ivanov@ikfia.ysn.ru}
\author{S.P. Knurenko}
\author{I.Ye. Sleptsov}
\affiliation{Institute of Cosmophysical Research and Aeronomy,
Yakutsk, 677891 Russia}

\begin{abstract}
The Yakutsk array observes the Cherenkov light emitted by UHECR in
atmosphere. Recently, an autonomous subarray is added consisting
of photomultipliers to measure the showers in the knee region. Our
aim is to analyze the combined data set in order to derive the
cosmic ray spectrum in the energy range as wide as possible using
the same technique.

The advantage of the air Cherenkov light measurement is the model
independent estimation of the EAS primary energy using the total
light flux emitted in the atmosphere. A set of the light lateral
distributions observed in the extended energy range is presented
also.
\end{abstract}

\pacs{95.85.R, 96.40} \keywords{Cosmic rays; Yakutsk array;
Cherenkov light; Energy spectrum}

\maketitle

\section{Introduction}
The Cherenkov light emitted in atmosphere by ultrarelativistic
electrons of extensive air showers (EAS) of cosmic rays (CRs)
carries the important information on the primary energy and shower
development in atmosphere. At the Yakutsk array air Cherenkov
light measurements are used to estimate the primary energy in the
model independent manner, and the lateral distribution form of the
light on the ground is used to infer the position of shower
maximum in atmosphere, $X_{max}$ ~\cite{Monograph}. In this paper
we have derived the energy spectrum of cosmic rays in the range
$\sim 1$ to $6\times 10^4$ PeV using the independent measurement
technique via air Cherenkov light and compared it with our
previous spectrum obtained with EAS charged particle measurements
as well as the results of other giant arrays.

\section{Air Cherenkov light measurements with the Yakutsk array}

\begin{figure}
\includegraphics[width=\columnwidth]{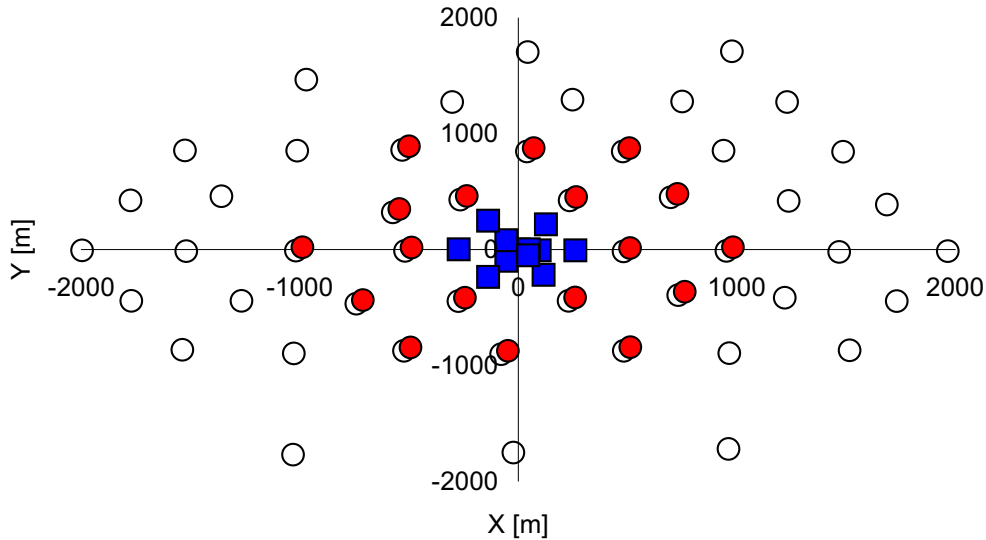}
\caption{\label{fig:map} The detector arrangement of the Yakutsk
array. A description of the points is given in the text.}
\end{figure}

During more than 30 years of lifetime the Yakutsk array is
detecting muonic, electronic and Cherenkov light components of
EAS. The actual detector arrangement of the array is shown
(Fig.~\ref{fig:map}). Open circles are charged particle detectors
given as the background here. Filled circles show the Cherenkov
light detectors - open photomultiplier tubes (PMTs) of 176
cm$^{2}$ and 530 cm$^{2}$ acceptance area, forming a medium
subarray. Filled squares indicate PMTs of autonomous subarray with
independent trigger. This subarray was added in 1995 with the aim
to study air showers in the energy range $10^{15}-10^{17}$ eV via
the Cherenkov light measurements.

During $\sim15000$ hours of observation in moonless nights there
were detected $\sim60000$ showers of energy above $6\times
10^{16}$ eV by the medium array. They were selected using the
double trigger condition:
i) three or more scintillator stations
have detected the charged particle density above the threshold 1
particle/m$^{2}$;
ii) The Cherenkov light intensities in three or
more PMTs are greater than $2.4\times 10^{5}/4.5\times 10^{5}$
photons/m$^{2}$ depending on the acceptance area of PMT.

The autonomous array data consist of $\sim 200000$ showers with
$E>1.2\times 10^{15}$ eV detected during $\sim 3200$ hours of
observation. The Cherenkov trigger condition only is used to
select these showers.

For absolute calibration of PMTs the Cherenkov radiation of
relativistic particles in distilled water was used
~\cite{Monograph}. The variations in cosmic ray intensity measured
with Cherenkov detectors itself are used to monitor the
atmospheric transparency ~\cite{Ocean}.

\section{Lateral distribution of the Cherenkov light}

\begin{figure}
\includegraphics[width=\columnwidth]{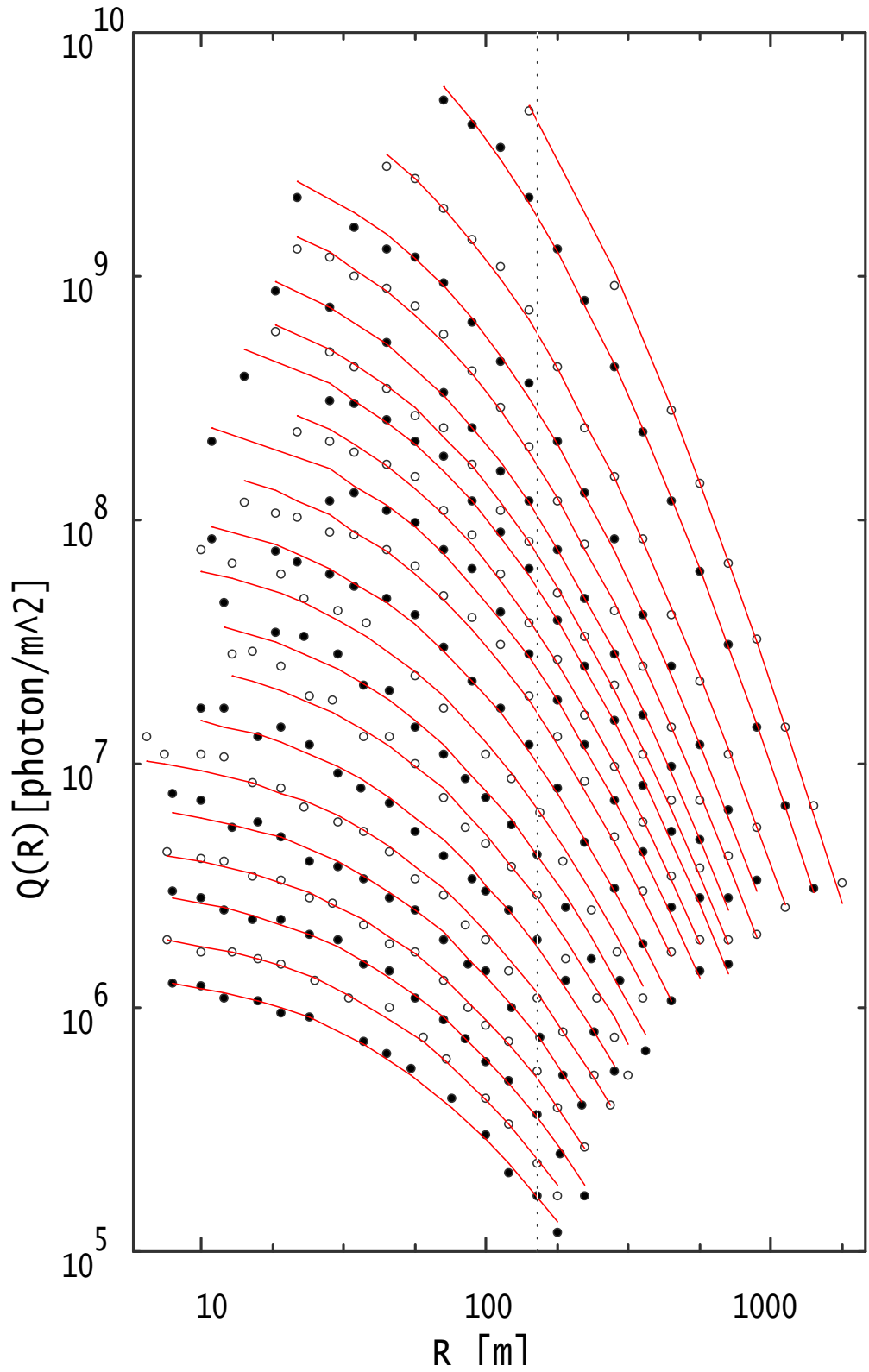}
\caption{\label{fig:LDF} Air Cherenkov light radial distributions.
Our experimental data are given by the points. The curves are
approximations along Equation~\ref{eq.ldf} for given Q$_{150}$.}
\end{figure}

We have summarized our lateral distribution function (LDF)
measurements of the Cherenkov light intensities in
Fig.~\ref{fig:LDF}. The data of both subarrays (zenith angle
$\theta<30^0$) are parameterized by the light intensity at 150 m
from the shower core, Q$_{150}$, the only core distance really
present in the shifting range of measurements when the primary
energy is rising from $\sim 10^{15}$ to 10$^{19}$ eV. The data are
consistent with the previous results of the Yakutsk array
concerning the Cherenkov light and can be described by the
suitable EAS model simulation ~\cite{Monograph}. There's seen no
abrupt change of LDF parameters, so we choose rather smooth
approximation curve to fit the experimental data in the whole
energy range:

\begin{equation}
Q(R)=Q_{150}\frac{(R_1+150)(R_2+R)^{1-b}}{(R_1+R)(R_2+150)^{1-b}},
\label{eq.ldf}
\end{equation}
where $R_1=60$ m; $R_2=200$ m;
$b=(1.14\pm0.06)+(0.30\pm0.02)\times \lg Q_{150}$

\section{Estimation of the primary energy of air showers}

The total flux, Q$_{tot}$, of the Cherenkov light emitted by the
shower in atmosphere is used to estimate the primary energy. The
detector design of the Yakutsk array is appropriate to measure
Q$_{tot}$ in the range above about $Q_{150}=10^7$
m$^{-2}$(Table~\ref{tab:table1}).

\begin{table}
\caption{\label{tab:table1}The fraction of Q$_{tot}$ actually
measured ($\Delta$).}
\begin{ruledtabular}
\begin{tabular}{ccccc}
 Q$_{150}$, m$^{-2}$ & 10$^6$ & 10$^7$ & 10$^8$ & 10$^9$ \\ \hline
 $\Delta,\%$ & 50 & 70 & 90 & 85 \\\end{tabular}
\end{ruledtabular}
\end{table}

In each Q$_{150}$ bin the LDF extrapolation formula~\ref{eq.ldf}
is used to calculate the total flux. In order to connect Q$_{tot}$
with the ionization loss, E$_i$, of the shower in atmosphere one
needs to know the atmospheric transparency factor for the
Cherenkov light and model predictions of the shower development.
We refer to ~\cite{Monograph} where this factor was given on the
basis of model calculations and transparency measurements in
Yakutsk:

\begin{equation}
E_i=\frac{2.18\times10^4 Q_{tot}}{0.37+1.1\times10^{-3}X_{max}},
\label{eq.Ei}
\end{equation}

This factor is nearly model-independent in a sense that the model
dependence is parameterized by $X_{max}$.

The fraction of the primary energy deposited in atmosphere can be
evaluated using the total number of electrons and muons measured
with the Yakutsk array, the attenuation length of the shower and
the mean energy of muons. The model calculation results are used
to estimate the energy of EAS nuclear active component and
neutrinos. The resulting apportioning of the primary energy
10$^{18}$ eV is given in Table~\ref{tab:table2}.

\begin{table}
\caption{\label{tab:table2}Estimation of the primary energy
portions gone with the shower components.}
\begin{ruledtabular}
\begin{tabular}{cc}
 Energy deposit channel &  The portion \\
                         &  of energy, $\%$ \\ \hline
 Ionization loss above sea level & 78 \\
 The total electromagnetic component energy & 87 \\
 Energy transferred to muons and neutrinos & 9 \\
 Energy carried by the nuclear active component & 4 \\
\end{tabular}
\end{ruledtabular}
\end{table}

We have used these values to derive the primary energy, $E$, of
showers with measured Q$_{150}$:

\begin{equation}
E=(1.07\pm 0.27)\times 10^{-5}Q_{150}^{(0.99\pm 0.02)}, PeV
\label{eq.E}
\end{equation}

\section{The energy spectrum of cosmic rays derived from air
Cherenkov light measurements}

\begin{figure}
\includegraphics[width=\columnwidth]{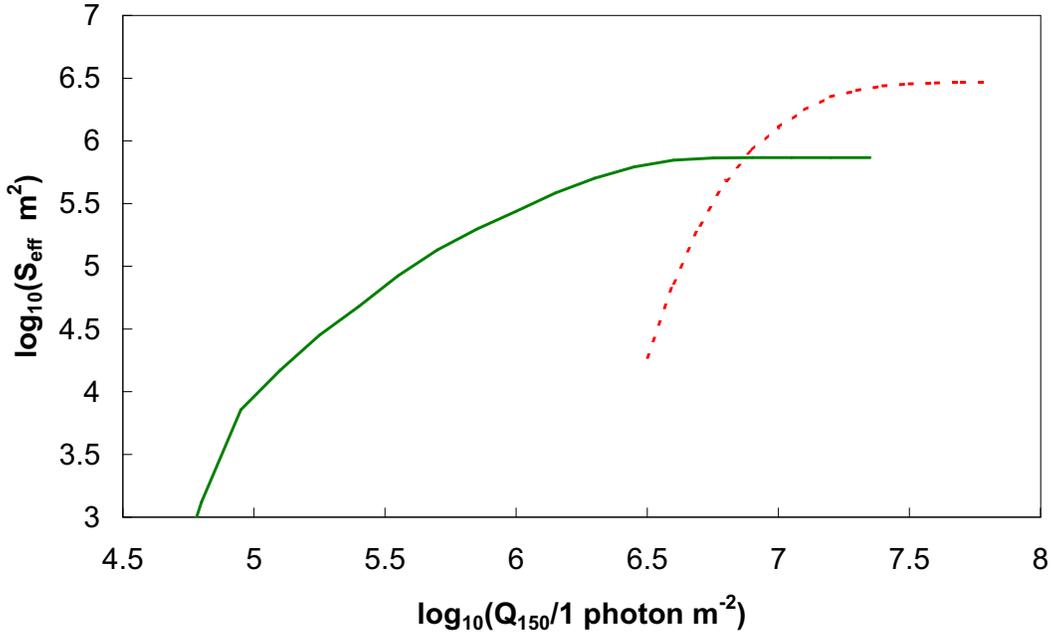}
\caption{\label{fig:Seff} Acceptance area of the Cherenkov
subarrays. The solid curve is for autonomous, and the dotted one
represents the medium subarray.}
\end{figure}

In order to evaluate the intensity of the primary flux we have
collected data during the observation periods with atmospheric
transparency better than 0.65, shower axes within area of
corresponding subarray, zenith angle $\theta<30^0$. Acceptance
area of the array, $S_{eff}$, depends on the primary energy and
zenith angle. We have calculated $S_{eff}$ as a function of
$Q_{150}$ averaged in zenith angle interval ($0^0,30^0$) for each
subarray (Fig.~\ref{fig:Seff}). Lateral distribution
fit~\ref{eq.ldf} is used along with instrumental and statistical
errors to model the trigger of each subarray with 100000 fake
showers. In the case of autonomous array the Cherenkov trigger is
simulated while in the medium case the double trigger for
scintillator and PMT signals is modeled. The shower data gathered
after the array was re-configured were used to work out the
spectrum. Namely, 1993-2001 for the medium, and 1995-2001 for
autonomous subarray.

\begin{figure}
\includegraphics[width=\columnwidth]{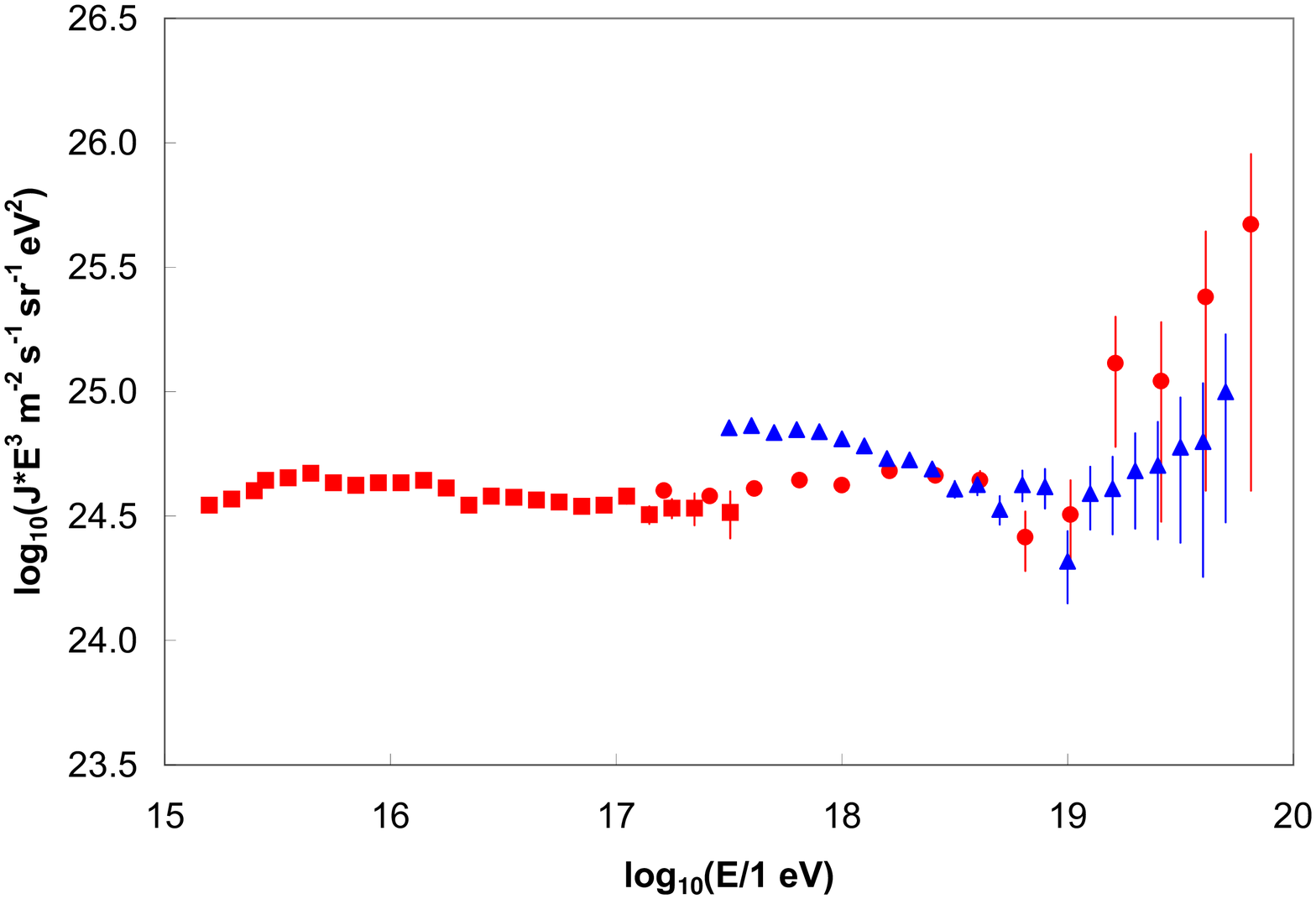}
\caption{\label{fig:SpectrOurs} The cosmic ray flux measured by
the air Cherenkov light (squares for autonomous and circles for
medium subarrays) and charged particle detectors (trigger-500,
triangles) of the Yakutsk array. Each data point represents the
differential flux scaled up by a factor of $E^3$. Error bars
represent statistical errors only.}
\end{figure}

The resulting differential all-particle spectrum of cosmic rays is
given (Fig.~\ref{fig:SpectrOurs},
Tables~\ref{tab:table3},~\ref{tab:table4}) together with the
Yakutsk array spectrum carried out by the charged particle
measurements with the trigger-500 m ~\cite{Kashiwa}. Spectra are
compatible above 2000 PeV within errors but are contradictory at
lower energies. We have to re-analyze our data near the threshold
energy of the scintillator trigger-500. Both methods confirm the
spectrum irregularity near $E\sim 10^4$ PeV (an 'ankle'), and the
autonomous subarray data is compatible with a 'knee' at $E\sim 3$
PeV. Our next spectrum obtained with trigger-1000 measurements of
charged particles ~\cite{Kashiwa} is in the energy range beyond
both shown spectra, out of reach of the Cherenkov subarrays,
therefore is not given here.

\begin{figure}
\includegraphics[width=\columnwidth]{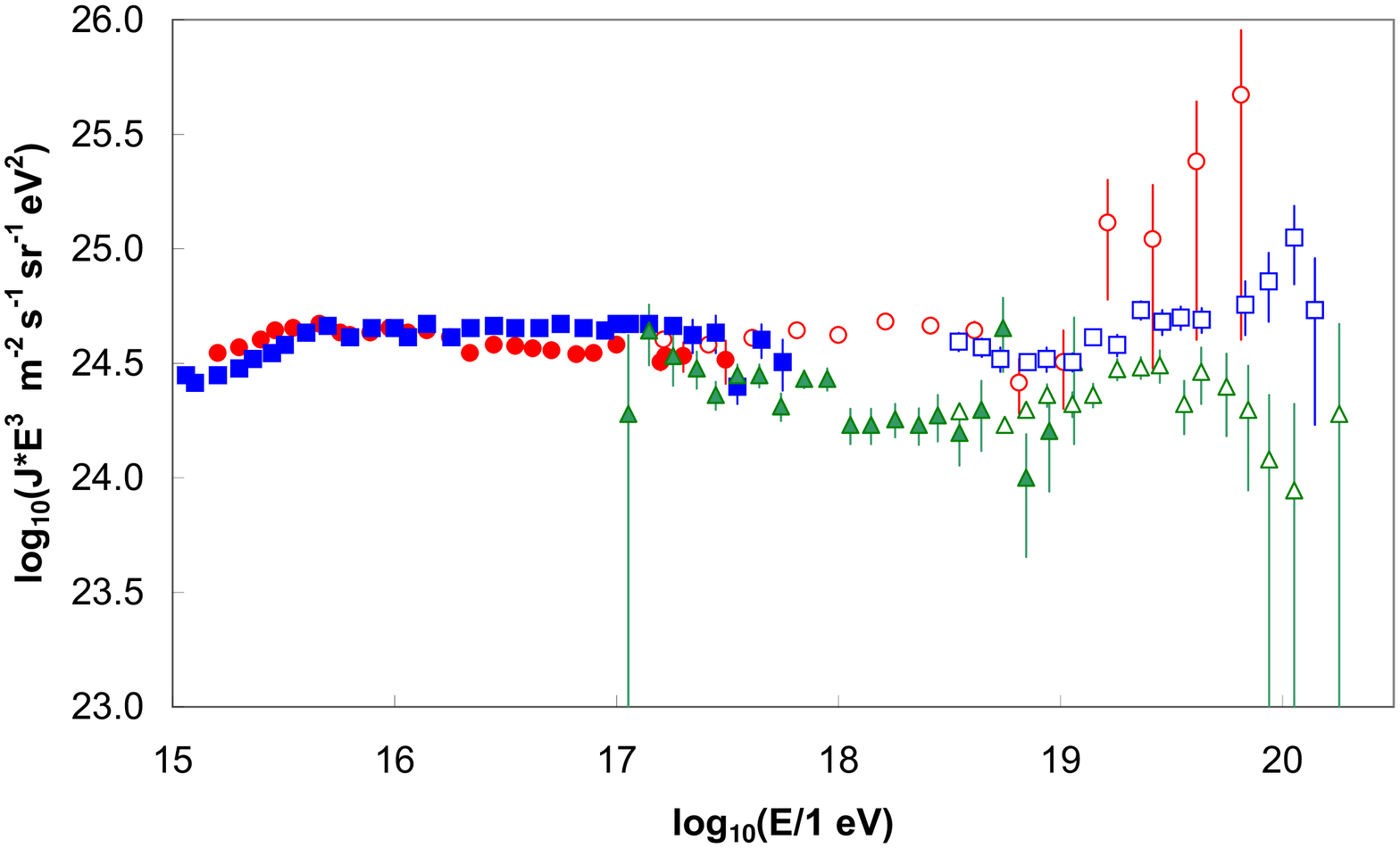}
\caption{\label{fig:SpectrAll} Our energy spectrum in comparison
with other experiments. The data are from HiRes-I (open
triangles), HiRes-II (filled triangles), Akeno (filled squares),
AGASA (open squares).  Our results are from the autonomous
subarray (filled circles) and medium subarray (open circles).}
\end{figure}

The autonomous subarray spectrum in the range 1 to 100 PeV was
given in comparison with other relevant results earlier
~\cite{Knurenko}. Comparison with results of other giant arrays
(Fig.~\ref{fig:SpectrAll}) shows a satisfactory agreement of the
flux with Akeno~\cite{Akeno} and AGASA~\cite{AGASA} data in the
energy range covered by the Cherenkov detectors of the Yakutsk
array. The data of HiRes-I and HiRes-II from the
preprint~\cite{HiRes} have the divergence with our spectrum which
can be explained, presumably, by the difference in the primary
energy estimation $\sim 30\%$. Re-evaluated energy spectrum of the
Haverah Park array ~\cite{HP} in the range $3\times 10^2$ PeV to
$4\times 10^3$ PeV coincides with the HiRes-II spectrum and isn't
given here.

\section{Conclusions}
The independent measurement technique using air Cherenkov light
detectors of the Yakutsk array enabled us to obtain the cosmic ray
energy spectrum in the range $\sim 1$ to $6\times 10^4$ PeV.
Although the PMTs arrangement of the array isn't convenient to
study air showers above $10^4$ PeV, there is an objective field of
activity below that border, wide enough. Our preliminary
'Cherenkov' spectrum contradicts previous 'scintillator' spectrum
around the threshold energy of the trigger-500. At the  higher
energies two spectra agree within errors. Both spectra exhibit an
'ankle' feature around $10^4$ PeV. The autonomous subarray data
confirm the 'knee' at $E\sim 3$ PeV. A comparison of our results
with the data of other giant EAS arrays shows the similarity of
spectra within $\sim 30\%$ uncertainty in the primary energy
estimation.

\begin{table}
\caption{\label{tab:table3}CR flux measured with the autonomous
subarray. Errors represent the statistical uncertainties.}
\begin{ruledtabular}
\begin{tabular}{rr}
Primary energy bin & Differential flux $J(E)$\\
 $\log_{10}(E/1 eV)$ &
{(m$^{-2}$\,sr$^{-1}$\,s$^{-1}$\,eV$^{-1}$)}
\\  \hline
 15.15 -- 15.25  & $ 8.54 \times 10^{-22}$ \\
 15.25 -- 15.35  & $ 4.63 \times 10^{-22}$ \\
 15.35 -- 15.45  & $ 2.56 \times 10^{-22}$ \\
 15.4  -- 15.5   & $ 1.80 \times 10^{-22}$ \\
 15.5  -- 15.6   & $ 1.05 \times 10^{-22}$ \\
 15.6  -- 15.7   & $ 4.83 \times 10^{-23}$ \\
 15.7  -- 15.8   & $ 2.32 \times 10^{-23}$ \\
 15.8  -- 15.9   & $ 1.68 \times 10^{-23}$ \\
 15.9  -- 16.0   & $ 9.06 \times 10^{-24}$ \\
 16.0  -- 16.1   & $ 2.83 \times 10^{-24}$ \\
 16.1  -- 16.2   & $ 1.60 \times 10^{-24}$ \\
 16.2  -- 16.3   & $ 7.27 \times 10^{-25}$ \\
 16.3  -- 16.4   & $ 3.33 \times 10^{-25}$ \\
 16.4  -- 16.5   & $ 1.73 \times 10^{-25}$ \\
 16.5  -- 16.6   & $ (8.77 \pm 0.42) \times 10^{-26}$ \\
 16.6  -- 16.7   & $ (4.95 \pm 0.23) \times 10^{-26}$ \\
 16.6  -- 16.7   & $ (4.95 \pm 0.23) \times 10^{-26}$ \\
 16.7  -- 16.8   & $ (2.71 \pm 0.14) \times 10^{-26}$ \\
 16.8  -- 16.9   & $ (1.20 \pm 0.07) \times 10^{-26}$ \\
 16.9  -- 17.0   & $ (7.10 \pm 0.43) \times 10^{-27}$ \\
 17.0  -- 17.1   & $ (3.80 \pm 0.25) \times 10^{-27}$ \\
 17.1  -- 17.2   & $ (8.11 \pm 0.66) \times 10^{-28}$ \\
 17.2  -- 17.3   & $ (7.43 \pm 0.66) \times 10^{-28}$ \\
 17.3  -- 17.4   & $ (4.25 \pm 0.63) \times 10^{-28}$ \\
 17.4  -- 17.6   & $ (1.10 \pm 0.23) \times 10^{-28}$ \\
\end{tabular}
\end{ruledtabular}
\end{table}

\begin{table}
\caption{\label{tab:table4}Cosmic ray flux measured with the
medium subarray detectors.}
\begin{ruledtabular}
\begin{tabular}{rr}
Primary energy bin & Differential flux $J(E)$\\
 $\log_{10}(E/1 eV)$ &
{(m$^{-2}$\,sr$^{-1}$\,s$^{-1}$\,eV$^{-1}$)}
\\  \hline
 17.1  -- 17.3   & $ 9.24 \times 10^{-28}$ \\
 17.3  -- 17.5   & $ 2.16 \times 10^{-28}$ \\
 17.5  -- 17.7   & $ 5.91 \times 10^{-29}$ \\
 17.7  -- 17.9   & $ 1.60 \times 10^{-29}$ \\
 17.9  -- 18.1   & $ 4.20 \times 10^{-30}$ \\
 18.1  -- 18.3   & $ 1.11 \times 10^{-30}$ \\
 18.3  -- 18.5   & $ (2.62 \pm 0.13) \times 10^{-31}$ \\
 18.5  -- 18.7   & $ (6.38 \pm 0.58) \times 10^{-32}$ \\
 18.7  -- 18.9   & $ (9.47 \pm 2.55) \times 10^{-33}$ \\
 18.9  -- 19.1   & $ (2.93 \pm 1.10) \times 10^{-33}$ \\
 19.1  -- 19.3   & $ (3.00 \pm 1.62) \times 10^{-33}$ \\
 19.3  -- 19.5   & $ (6.26 \pm 4.55) \times 10^{-34}$ \\
 19.5  -- 19.7   & $ (3.48 \pm 2.90) \times 10^{-34}$ \\
 19.7  -- 19.9   & $ (1.71 \pm 1.57) \times 10^{-34}$ \\
\end{tabular}
\end{ruledtabular}
\end{table}

\begin{acknowledgments}
The authors are grateful to Russian ministry of Science and
Technology for the support of the Yakutsk array under the program
\#01-30 'Unique research facilities in Russia'. Our special thanks
to the Yakutsk group members who made the essential contributions
to data acquisition and analysis. This work is supported by RFBR
grants \#00-15-96787, \#02-02-16380.
\end{acknowledgments}


\end{document}